\pdfoutput=1

\documentclass[final,3p,times]{elsarticle}

\usepackage{graphics}
\usepackage{graphicx}
\usepackage{epsfig}
\usepackage{amssymb}
\usepackage{float}
\usepackage{color}
\journal{Adv. in High Energy Phys.}

\usepackage[normalem]{ulem}
\usepackage{color}
\definecolor{blue}{rgb}{0,0,1}
\definecolor{red}{rgb}{1,0,0}

\begin{document}
\begin{frontmatter}
\title{High efficiency gaseous tracking detector for cosmic muon radiography}

%%%%%%%%%%%%%%%%%%%%%%%%%%%%%%%%%%%%%%%%%%%%%%%%%%%%%%%%%%%%%%%%%%%%%%%%

\author[1]{Dezs\H o Varga}
\ead{varga.dezso@wigner.mta.hu}
\author[2]{G\'abor Nyitrai}
\author[1]{Gerg\H{o} Hamar}
\author[1,3]{L\'aszl\'o Ol\'ah}
\ead{olah.laszlo@wigner.mta.hu}

\address[1]{Wigner Research Centre for Physics of the Hungarian Academy of Sciences, 29-33 Konkoly-Thege Mikl\'os Str. H-1121 Budapest, Hungary}
\address[2]{Budapest University of Technology and Economics, 3 M\H{u}egyetem rkp., H-1111 Budapest, Hungary}
\address[3]{E\"otv\"os Lor\'and University, 1/A P\'azm\'any P. s\'et\'any, H-1117 Budapest, Hungary}

%%%%%%%%%%%%%%%%%%%%%%%%%%%%%%%%%%%%%%%%%%%%%%%%%%%%%%%%%%%%%%%%%%%%%%%%%%%%%%%%%%%%%%%%%%%%%%%%%%%%%%%%%%%%%%%%%%%%%%%%%%%%%%%%%%%%%%%%%%%%%%%%%%%%%%%%%%%%%%%%%%%%%%%%%%%%%%%%%%%%%%%%%%%%%%%%%%%%%%%%%%%%%%%%%%%%%%%%%%%%%%%%%%%%%%%%%%%%%%%%%%%%%%%%%%%%%%%%
%%%%%%%%%%%%%%%%%%%%%%%%%%%%%%%%%%%%%%%%%%%%%%%%%%%%%%%%%%%%%%%%%%%%%%%%%%%%%%%%%%%%%%%%%%%%%%%%%%%%%%%%%%%%%%%%%%%%%%%%%%%%%%%%%%%%%%%%%%%%%%%%%%%%%%%%%%%%%%%%%%%%%%%%%%%%%%%%%%%%%%%%%%%%%%%%%%%%%%%%%%%%

\begin{abstract}
\noindent
A tracking detector system has been constructed with an innovative approach to the classical multi-wire proportional chamber concept, using contemporary technologies. The detectors, covering an area of 0.58 square meters each, are optimized for the application of muon radiography. The main features are high ($>$99.5\%) and uniform detection efficiency, 9 mm FWHM position resolution, filling gas consumption below 2 liters per hour for the non toxic, non flammable argon and carbon dioxide mixture. These parameters, along with the simplicity of the construction and the tolerance for mechanical effects, make the detectors to be a viable option for a large area muography observation system. 

\end{abstract}

\begin{keyword}
gaseous tracking detectors \sep cosmic muon detection \sep muon radiography
     
\end{keyword}

\end{frontmatter}


%\linenumbers

%%%%%%%%%%%%%%%%%%%%%%%%%
%	INTRODUCTION
%%%%%%%%%%%%%%%%%%%%%%%%%

\section{Introduction}
\label{sec:intro}

Cosmic muons are highly penetrating particles, which are created by interaction of high energy cosmic particles impinging on the upper atmosphere of Earth. Muons are decay products of pions, which make up the majority of the particles reaching the Earth surface. Production mechanisms and interaction properties are extensively discussed in the literature~\cite{grieder:2001}. The mean energy of muons is around 2 GeV, and such an average muon would be able to cross 3 -- 4 meters of rock. This considerable penetrating power gives rise to the practical applicability of cosmic muon radiography, or ``muography'' in short.

There is a broad range of applications of muography. Very large objects can be imaged by measuring muon absorption, such as in the classical measurement by Alvarez {\it et al}~\cite{alvarez:1970}. Notable, high impact examples are volcano imaging~\cite{nagamine:1995, tanaka:2007, tanaka:2014} and mapping the core of nuclear reactors~\cite{borozdin:2012}. Underground measurements have smaller background due to the natural shielding, and can be exploited during ore mining~\cite{canada_mining} or natural cave search~\cite{caffau:1997, mt_nima:2012, mt_geo:2012, olah:2013}. If the object of interest is not exceedingly large, measuring the muon scattering~\cite{borozdin:2003, schultz:2004, pesente:2009, gnanvo:2011} can provide information about the internal structure.

When it comes to particle detectors for muon radiography, one of the key considerations is the flux of the signal particles. On the surface, the rate is in the order of one hundred particles per square meter per second -- a relatively low value compared to high energy physics (HEP) experimentation. On the other hand, if a large object is to be imaged, the flux of the muons can be drastically reduced further~\cite{nagamine:1995}: this implies that large size detectors are mandatory for most meaningful measurements. An additional important point is the suppression of background against the relevant (transmitted) muon signal: this background turns out to be mostly low energy particles from various sources~\cite{nishijama:2014, nishijama:2016}.

From the above it follows that the requirements for muography detectors differ from those generally expected from high energy physics detectors (for the latter, see textbooks such as Ref.~\cite{blumrolandi:1993}). The key points may be summarized as follows:

\begin{itemize}

\item{Large size at sensible cost. Due to the low and fixed cosmic rate, the statistics becomes the ultimate limiting factor of the measurements on large objects.}

\item{Long term operation under ambient outdoor environmental conditions. Considerable temperature, pressure and humidity variations need to be tolerated.}

\item{Tracking capability. In typical HEP conditions, the detector size is usually constrained by available space. On the contrary, most applications of muography require a specific angular resolution, which by increasing the measurable track length, allows for weaker position resolution. However, clear identification of the straight muon tracks is mandatory to suppress low energy background.}

\item{Rate capability. The cosmic particle rate is low on the Earth surface, therefore none of the issues apply which characterize ``high rate'' detectors in HEP. Eliminating those design elements which are meant to improve rate capability potentially results in considerable simplification of the structure.}

\item{Sustainable and safe operation. Given the long continuous measurement times, low maintenance and low running costs are desirable, whereas outdoor installation requires safe and environmentally friendly operation.}

%\item{Mobility, portability, modularity. ???? }

\end{itemize}

Various detector technologies have been considered for muon radiography. The three notable groups are scintillators, emulsion and gaseous detectors. So far scintillator-based trackers proved to be the most successful, applied for vulcanology~\cite{tanaka:2003, lesparre:2012, carbone:2013} and mining~\cite{canada_mining}. Emulsions present very high position resolution~\cite{morishima:2015}, and hence the detector layers require small space~\cite{tanaka:2007b, tioukov:2014, nishijama:2014}. The main drawback of this technology is that the information can be made available only after the film development, making real time measurements impossible. Gaseous detectors have the possibility to reach relatively large surfaces at reasonable cost, and therefore these tend to be the choice for the outer layers of most HEP experiments. Various groups proposed viable solutions with Resistive Plate Chambers~\cite{carloganu:2013}, drift chambers~\cite{schultz:2004, pesente:2009} and GEMs~\cite{gnanvo:2011}. %Drift tubes have been commercialized for high density contraband detection~\cite{}. 

Our objective was to conceive a tracking detector system which is optimized for muography from the very beginning of the design. Based on our own experience with various gaseous detectors, an innovative approach to the classical Multi Wire Proportional Chambers (MWPC-s) \cite{blumrolandi:1993} was chosen. In smaller sizes, MWPC-s with simplified geometry may be used for educational purposes~\cite{mwpc}. 

In order to address the ``muography-requirements'' above, design features include consistent material choice (Section \ref{sec:construct}) for improved thermal behaviour, an internal support structure with causes little efficiency loss (Section \ref{sec:deadzone}) and improves mechanical stability (\ref{sec:mech_stab}). As working gas, the mixture of argon and carbon dioxide was studied in detail (Section \ref{sec:gasstudy}), being the most obvious choice from cost, environmental and safety considerations.

The paper presents the detector construction, followed by relevant performance studies, for a sensitive area of about 0.58 ${\rm m}^2$. The key guiding principle was the simplification of the design, without compromise on detection efficiency and tracking capability: our studies below attempt to give a comprehensive performance report.

%Considering rate cabability as an example, all those elements can be eliminated from the design which are meant to deal with high rate (high in terms of high energy physics experiments). 

\section{Construction}
\label{sec:construct}

\begin{figure}[h!]
\begin{center}
\includegraphics[ angle=0, width=0.7\textwidth ]{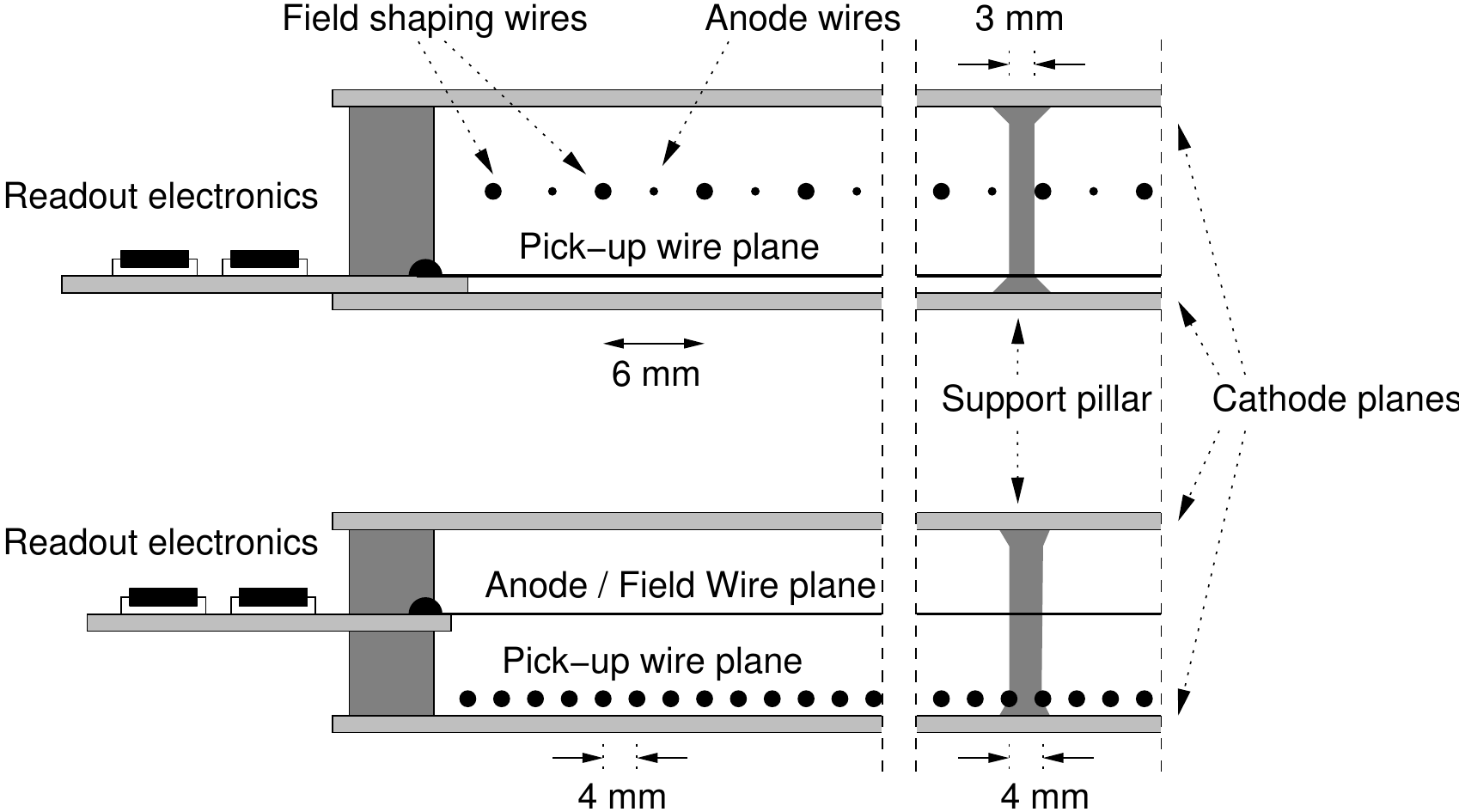}
\caption{Cross section of the detector close to the side showing wire fixing, in both directions (perpendicularly to, and parallel with the anode wires). The support pillar positioning between the wires are also indicated.}
\label{fig:xsect}
\end{center}
\end{figure}

The detector uses a relatively large, 22 mm gas gap between two grounded cathode planes. The gas amplification takes place on an anode (sense, SW) wire plane nearly half way between the cathodes. In the same plane, field shaping wires (FW) are fixed as well parallel to the anode wires, as shown in Figure \ref{fig:xsect}. There is an other wire plane, for the so called ``pick-up wires'' (PW), which run perpendicularly to the anode wires. The concept is similar to the CCC case~\cite{ccc:2011, ccc:2013}: the anode wires are used as trigger and amplitude measurement, whereas the field shaping wires and the pick-up wires give the position information. The pick-up wire plane is 2 mm above one of the cathodes, as shown in Fig.~\ref{fig:xsect}. The advantage of using such pick-up wires is cost consideration and simplification of the structure: an alternative would be using printed circuit boards (PCB-s) with strips, however PCB-s at large sizes tend to be expensive and their surfaces are not sufficiently flat. The anode wires for the specific detectors presented in this paper are 22 and 25 micrometers thick Au-plated tungsten, whereas both FW and PW are brass wires of 100 micrometers diameter. The two different anode wire diameters presented no practical difference except for a slight change in operational voltage.

\begin{figure}[h!]
\begin{center}
\includegraphics[ angle=0, width=0.4\textwidth ]{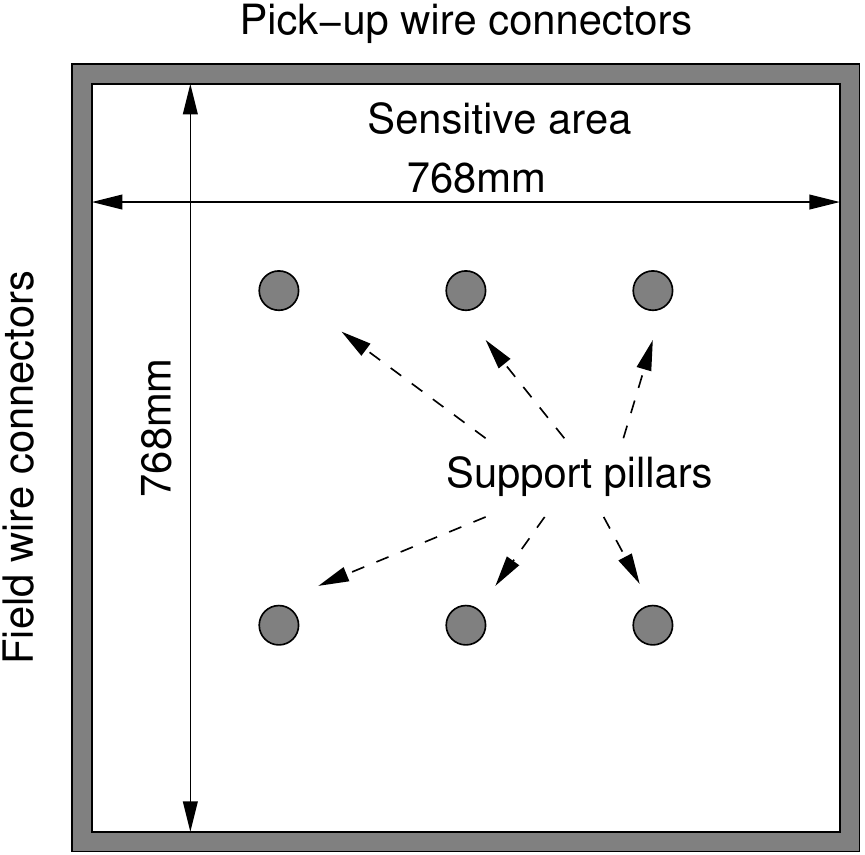}
\caption{Schematic representation of the detector sensitive area. Field and pick-up wire connectors cover one sides each. The approximate positions of the internal support pillars (in our case 6 of them) is also indicated (not to scale).}
\label{fig:80cm_outline}
\end{center}
\end{figure}

In one detector panel, there are 64 field shaping wires (hence 63 sense wires), with wire spacing of 12 mm, covering a 768 mm long region. The pick-up wires cover the same length with 4 mm spacing (hence there are 192 of them), which are connected in groups of 3 (that is, each of the 64 pick-up wire groups are 12 mm wide). All the wires are approximately 770 mm long. The front view outline is shown in Fig.~\ref{fig:80cm_outline}.

From the electronic connection point of view, the 64 field wires and the 64 pick-up wire groups are directly connected to the readout (see in Sec.~\ref{sec:daq}), whereas the 63 sense wires are all interconnected. From the sense wire signal, two output signals are formed after pre-amplification and shaping: one is a TTL-compatible discriminated trigger signal, and the other is a direct analog output for amplitude measurement.

The detector is constructed out of fiberglass (glass fiber reinforced epoxy). Specifically the sidewalls are G-10, whereas the cathode planes and other printed circuit components (soldering points, electronics support) are standard FR-4. This way, all parts of the detector feature a consistent thermal behavior. The cathode planes are 1.5 mm thick FR-4 layers, fully covered with copper on both sides. The complete detector is glued with epoxy: the units are thus not repairable in case of broken wires (however, no such case was observed so far after a year of operation). The total weight of a detector panel is about 6 kg including electronics.

Given the relatively large area, the distance between the cathode planes need to be kept at the nominal 22 mm with reasonable precision. In order to achieve this, internal support pillars were installed in a particular way between the wires, as shown in Fig.~\ref{fig:xsect}. The pillars have a rectangular 3mm by 4mm cross section around the anode wire plane. These pillars, made out of insulating material (PLA) using 3D printing technology, do not touch any of the field and sense wires. The pillars were placed closer to the field wires, with around 0.5 mm clearance, which leaves approximately 2.5 mm distance from the sense wire. One can expect that near a support pillar, the field lines still terminate on the sense wires, therefore the detector remains sensitive to the traversing particles -- a feature which proves to be correct according to the studies presented in Sec.~\ref{sec:deadzone}. For our particular detectors, 6 such pillars have been installed, according to Fig.~\ref{fig:80cm_outline}. The possibility of installing many of such pillars without introducing a large dead zone in the detector is of key importance for mechanical stability.

The detectors were operated with mixture of argon and carbon dioxide, in proportion of 82:18, at various flow rates ranging from 1 to 10 liters per hour. Detailed studies on gas behavior is presented in Section \ref{sec:gasstudy}.

\section{Tracking detector design}
\label{sec:tracking_design}

A tracking system has been set up, using six individual detector layers, arranged each with 20 cm spacing (that is, 1 m distance between the first and the last layer). The detectors were mounted vertically, standing on the side without connectors (bottom side in Fig.~\ref{fig:80cm_outline}), that is, the anode wires were horizontal. The detectors are counted from 1 to 6 for the rest of the paper. For most of the measurements, the detectors shared the same anode high voltage. The gas line was connecting the layers in series: the gas entered to detector 1, and was vented to the atmosphere from detector 6.  

\begin{figure}[h!]
\begin{center}
\includegraphics[ angle=0, width=0.5\textwidth ]{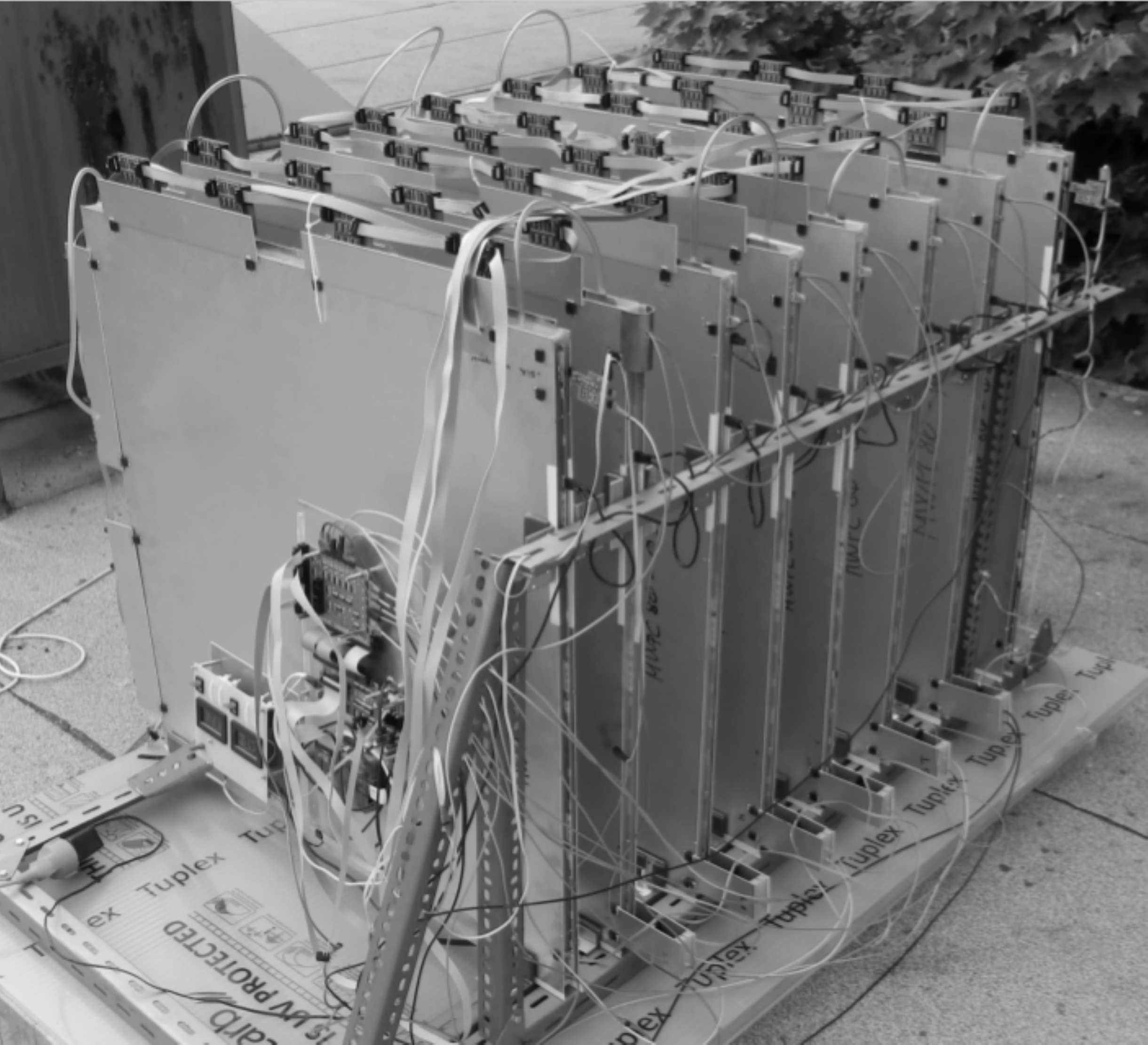}
\caption{Image of the tracking system setup on the roof installed with vertical detectors. A light cover box was used in addition to give basic environmental protection.}
\label{fig:mt}
\end{center}
\end{figure}

The setting with vertical detectors resembles most to the application in volcanology, where the object is imaged from the ``side'', and where, because of the reduced muon flux close to the horizon, the system is most sensitive to background. Such dedicated issues are however outside of the scope of the present paper: for the tracking system we did not use absorbers, which are otherwise mandatory for such arrangements~\cite{tanaka:2014}, and will be subject to later detailed studies. The image of the actual system with vertical detectors is shown in Fig.~\ref{fig:mt}, including two additional non-functional layers.

For various performance studies, we applied a similar arrangement, but with horizontal detectors inside the laboratory: that is, capturing the most abundant nearly vertical muons, high statistics mappings of the detector parameters became possible.

The six detector layers are sufficient to identify clear particle tracks, and allow for the actual detector performance studies. We have used tracking algorithms from our own earlier systems \cite{mt_geo:2012}. From the tracking information one can tune the detector geometry (by requiring straight tracks on average), which is needed for a more precise track finding. The tracking system has shown a consistently excellent performance, enabling a reliable characterization of the individual detector layers using cosmic muons.

\section{Readout and data acquisition}
\label{sec:daq}

The detector readout was adapted from our own earlier system~\cite{olah:2013}, with a single Raspberry Pi microcomputer as the core of the data acquisition~\cite{olah:2015}. An event sequence starts with the arrival of the trigger signals: if three coincident triggers appear (from any of the 6 layers), the readout will wait for a pre-defined delay time, and then issue a ``master'' trigger pulse. At the instance of this master trigger, the position information is recorded from discriminated signals of all the channels of the field wires and the pick-up wires. Also the master trigger time defines when the sense wire amplitude measurement, via 10 bit ADC-s, takes place.

The total data from the tracking system is moderate: there are 6$\times$2$\times$64 = 768 bits from the position information, and including the converted ADC values as well as recording the trigger pattern, the raw data is still around 100 Byte per event. This explains why a single Raspberry Pi can handle the DAQ task. 

%The readout time for a complete single event was 0.8 ms.

The input signals from the wires are protected from external electronics noise as much as possible, and since the grounded cathodes act as a sort of Faraday cage, a reasonable noise performance has been observed. The trigger level from each of the layers are set approximately to $2 \times 10^5$ electrons (32 fC) on input; for the field shaping and pick-up wires the discrimination threshold was also set to $2 \times 10^5$ electrons (measured as charge on the sense wires). 

\section{Efficiency of cosmic muon detection}
\label{sec:efficiency}

A key figure of merit of a detector is the tracking detection efficiency, that is, the probability that a valid particle hit (position) is actually detected. This we define in the following way for the rest of the paper: considering a specific detector layer, a particle trajectory is formed from the 5 detectors excluding the one under study. If this 5-point pattern is a unique, straight track from both directions (see Section \ref{sec:resolution}), one checks if there is a hit at the expected position (within 5 sigma of the position resolution) in the detector under study. The probability of finding the hit is defined as the ``tracking efficiency''. In a similar way, one can check if the specific detector layer provided a trigger signal for the event: this defines is the ``trigger efficiency''. Trigger and tracking efficiencies are found to be well above 90\% for the operational conditions.

%, which means that the biases caused by requiring the 5-point tracks are small.

\begin{figure}[h!]
\begin{center}
\includegraphics[ angle=0, width=0.7\textwidth ]{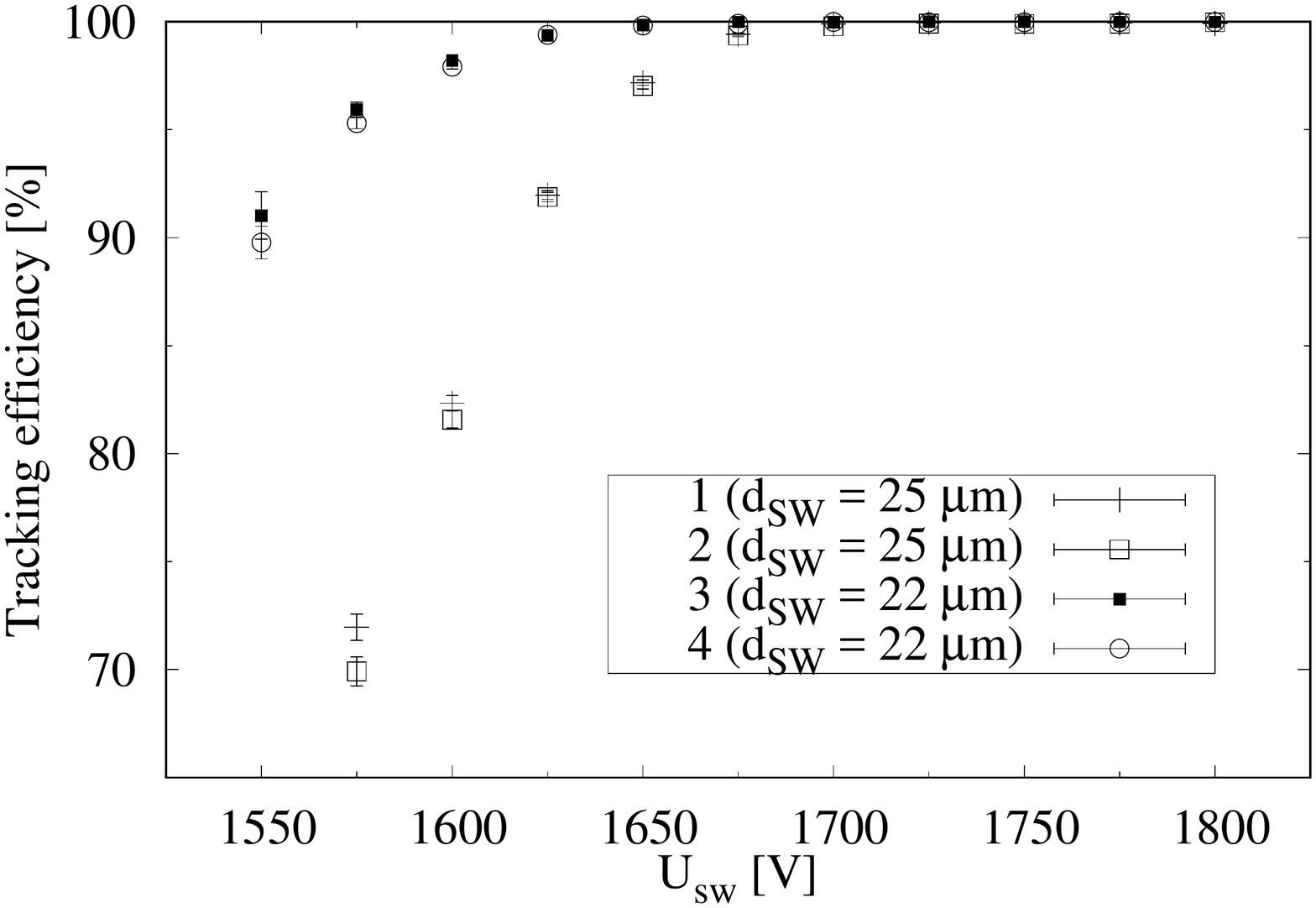}
\caption{Tracking efficiency of the detectors 1 - 4 as a function of the anode high voltage. A very clear plateau is observed, which extends from 1700 V to 1800 V. The detectors were built with 22 and 25 $\mu$m diameter anode wires as indicated.}
\label{fig:eff_hvscan}
\end{center}
\end{figure}

First, the tracking and trigger efficiency has been determined as a function of the sense wire voltage. A clear plateau is observed, shown in Fig.~\ref{fig:eff_hvscan}. The efficiency reaches around 95\% ($>$99\%) at 1650 V (1700 V) for the detectors with 25 micron anode wires. The plateau continues to 1800 V, where stable operation has still been observed.

\begin{figure}[h!]
\begin{center}
\includegraphics[ angle=0, width=0.7\textwidth ]{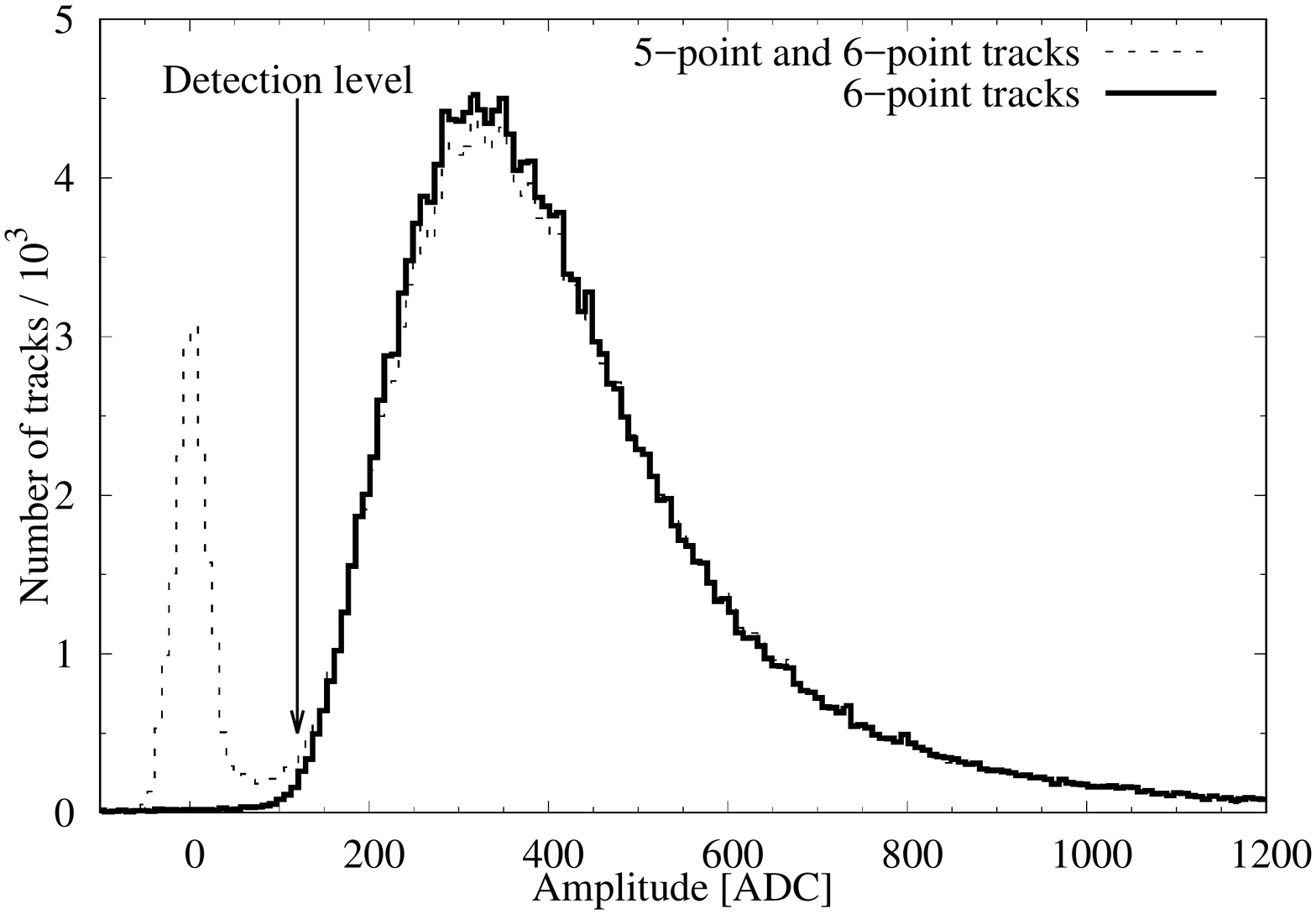}
\caption{Signal amplitude spectra in the first layer for tracks with at least 5 points ({\it dashed lines}), and for tracks with 6 points ({\it continuous line}) with the approximate detection level of 120 ADC units. The peak around zero amplitude corresponds to those tracks which miss the first layer.}
\label{fig:landau}
\end{center}
\end{figure}

The efficiency values shown in Fig.~\ref{fig:eff_hvscan} are consistent with the expectation from the signal amplitude spectrum, shown in Figure \ref{fig:landau} for a sense wire voltage of 1700 V for the first tracking layer. The spectrum of all tracks with at least 5 points are overlayed with the spectrum of 6-point tracks. The difference between these two sets, in case of full efficiency, is just those tracks which miss the first layer and hence result in a 5-point track. The approximate detection level is 120 ADC units which has also been indicated with an arrow. The horizontal scale of Fig.~\ref{fig:landau} is expressed in ADC units, which have been calibrated by direct charge injection: 1 ADC unit corresponds to 1.8 $\times 10^3$e, or 0.3 fC. 

The true amplification gain, that is, the multiplication factor of the avalanches, has been estimated measuring the mean amplitude on the sense wires. Taking into account that the mean number of electrons over the 2 cm of the drift gap is around 200 in the applied gas mixture~\cite{blumrolandi:1993}, one finds that the detectors with 25 micron anode wires at 1700 V (full efficiency) has a gain of about 7 $\times 10^3$, a conveniently low figure.

\begin{figure}[h!]
\begin{center}
\includegraphics[ angle=0, width=0.7\textwidth ]{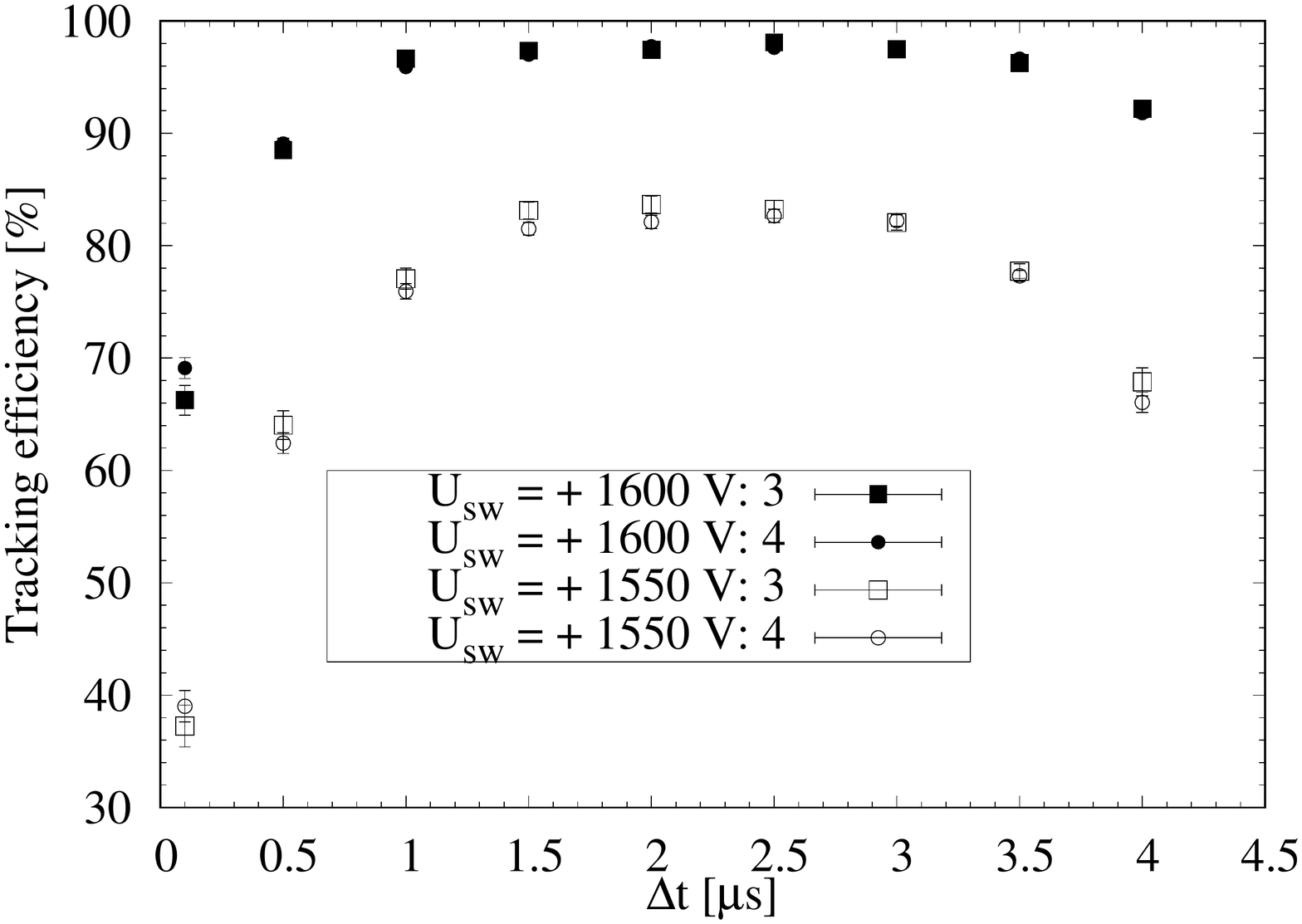}
\caption{Signal timing curves: the tracking efficiency of the two inner layers as a function of the delay relative to the arrival of the coincident trigger (mostly from the four other detectors, operated using high efficiency settings, see text). The measurement is consistent with the observed signal peaking time of 2 $\mu$s, the value used as the ``standard'' delay setting.}
\label{fig:eff_timing}
\end{center}
\end{figure}

The trigger signal is created at the point when the input signal reaches the trigger threshold. There is a certain time delay needed after this to reach the highest amplitude on the field wire and pick-up wire discriminators, and hence to reach optimal tracking efficiency. The tracking efficiency as a function of this trigger delay is shown in Figure \ref{fig:eff_timing}, with a maximum around 2-3 $\mu s$ consistent with the peaking of the signal. The data has been taken keeping detectors 1,2,5,6 at 1700 V and reducing the voltage on the middle detectors, 3 and 4, to the values indicated (1600 V and 1550 V). This procedure ensures reasonably high trigger efficiency (from four detectors) and a measurable efficiency (that is, well below 100\%) for the inner layers. The setting for all other measurements use a fixed 2 $\mu$s delay.

\section{Position resolution}
\label{sec:resolution}

\begin{figure}[h!]
\begin{center}
\includegraphics[ angle=0, width=0.5\textwidth ]{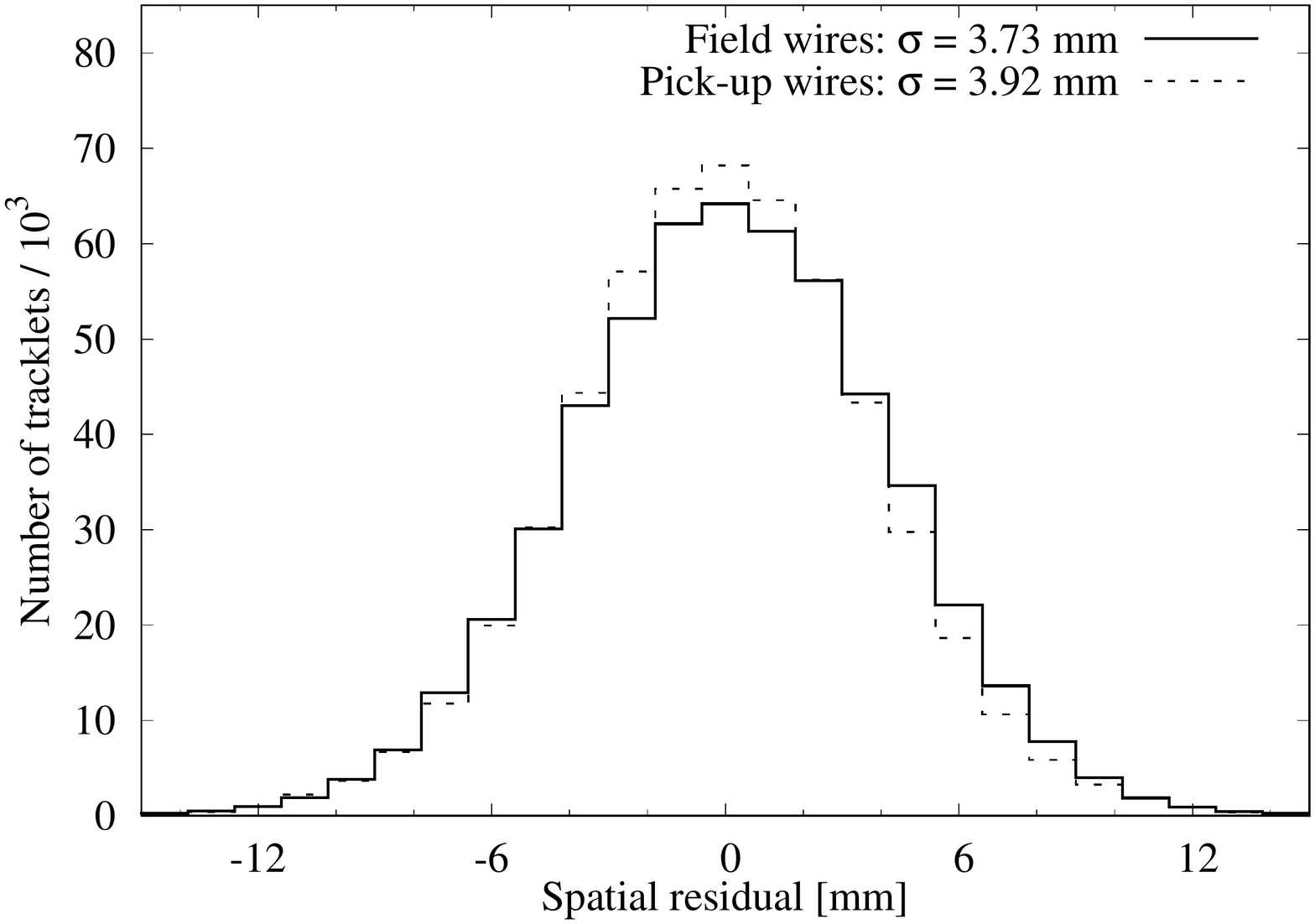}
\caption{The position resolution of the detector, measured as the distance between the observed hit position and the fitted track intersection (well defined trajectory identified in the detector, excluding the layer under study). The resolution in both directions is found to be below 4 mm.}
\label{fig:pos_res}
\end{center}
\end{figure}

With the 6-layer tracking detector setup, the position resolution of the layers can be quantified. Figure \ref{fig:pos_res} shows the distance between the measured hit in one of the middle layers and the extrapolation from the 5-point track (that is, the fitted track excluding the specific layer). One finds that the position resolution (RMS) is $\sigma_{FW}$=3.73 mm in the field wire, and $\sigma_{PW}$=3.92 mm in the pick-up wire direction, that is, around 9 mm FWHM. This is roughly consistent with the 12 mm spacing expectation ($12{\rm mm} / \sqrt{12} = 3.5 {\rm mm}$). The tracking algorithm incorporates this position resolution to find a set of points on 5 or 6 of the chambers which form a straight line from both directions.

%\section{Edge effects and dead zones}
\section{Dead zones due to support pillars}
\label{sec:deadzone}

\begin{figure}[h!]
\begin{center}
\includegraphics[ angle=0, width=0.5\textwidth ]{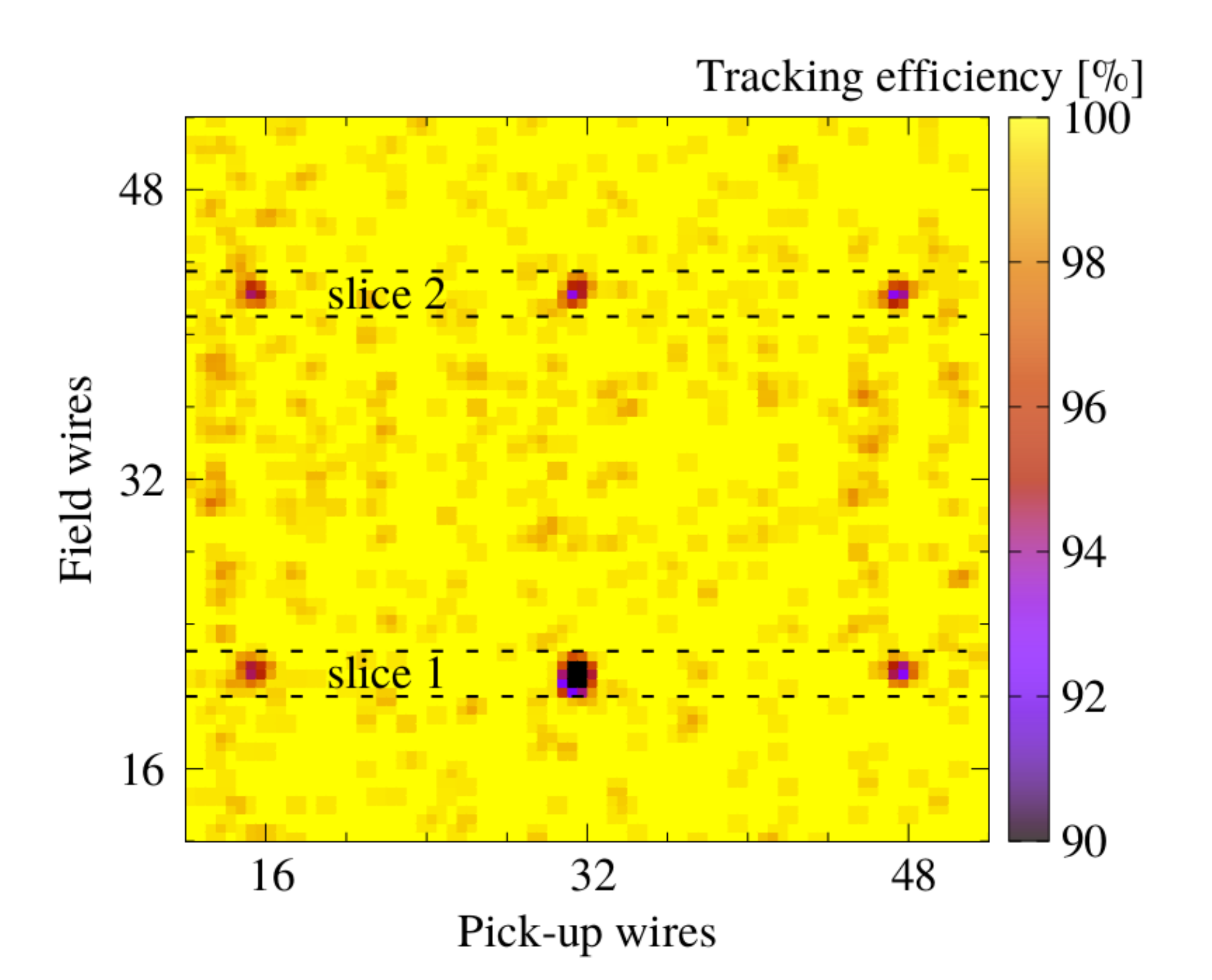}
\caption{(Color online) Efficiency map covering the region of the support pillars. The reduction is small and localized (note the zero suppressed scale).}
\label{fig:pillar2d}
\end{center}
\end{figure}

\begin{figure}[h!]
\begin{center}
\includegraphics[ angle=0, width=0.7\textwidth ]{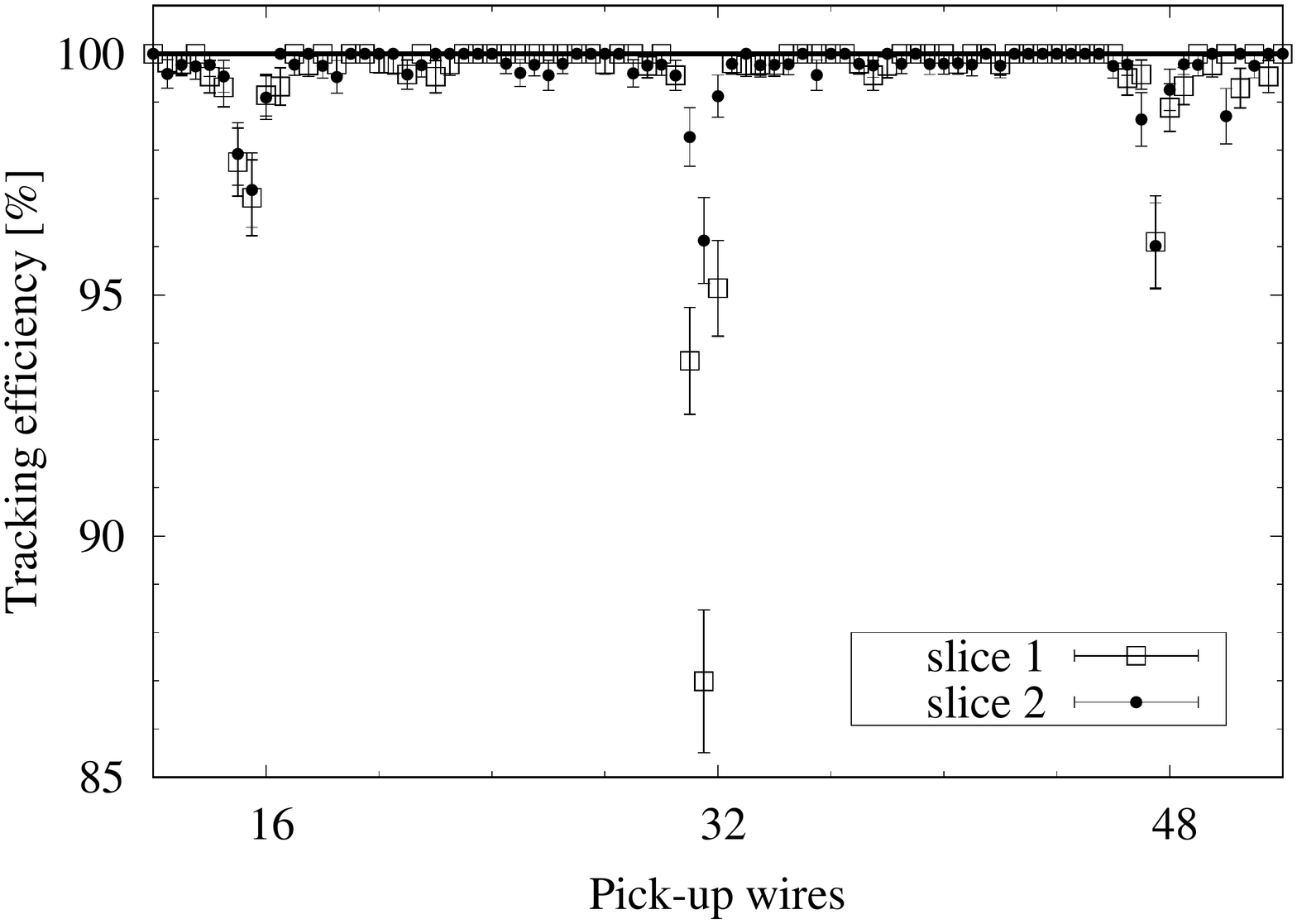}
\caption{Efficiency drop around the support pillars averaged over the 3 cm wide slice indicated with the dashed lines in Figure \ref{fig:pillar2d} . }
\label{fig:pillar1d}
\end{center}
\end{figure}

The detector, as seen above, is highly efficient except for the support pillars, which deserve a closer look. In the tracking system configuration with horizontal detectors, the local efficiency has been determined, as shown in Figure \ref{fig:pillar2d}. For this measurement, nearly vertical tracks were selected, with less than 200 mrad relative to the zenith. The pixel size in this map is one half of the readout unit (6 mm), and one immediately notes the surprisingly small reduction. (The measured efficiency does not go to zero due to the fact that the pillars are smaller than the tracking detector position resolution). Taking a 3 cm wide strip over the pillars (indicated in the Figure by two dashed lines), the mean efficiency is indicated in Figure \ref{fig:pillar1d}. To first order, one can conclude that the detector is \emph{sensitive for any muons which enters the gas volume around the pillars}, or in other words, the presence of the support pillars does not distort the field structure to the extent to considerably reduce efficiency. There seems to be some fluctuation on the apparent loss caused by the pillars, which may have to do with the details of the actual positioning.

%Side edge effects???? Gain map at the edge????

\section{Mechanical stability}
\label{sec:mech_stab}

The mechanical stability and required tolerances of the detector is a key question for practical outdoor applications. In the studies below, some of the mechanical effects on the detectors are quantified in relation to its performance. Despite the simplicity of the structure, it turned out that most of the relevant mechanical variations (e.g. difference between vertical or horizontal position) give insignificant changes, therefore we decided to expose the detectors to considerable mechanical effects, outside of the normally expected conditions.

\begin{figure}[h!]
\begin{center}
\includegraphics[ angle=0, width=0.8\textwidth ]{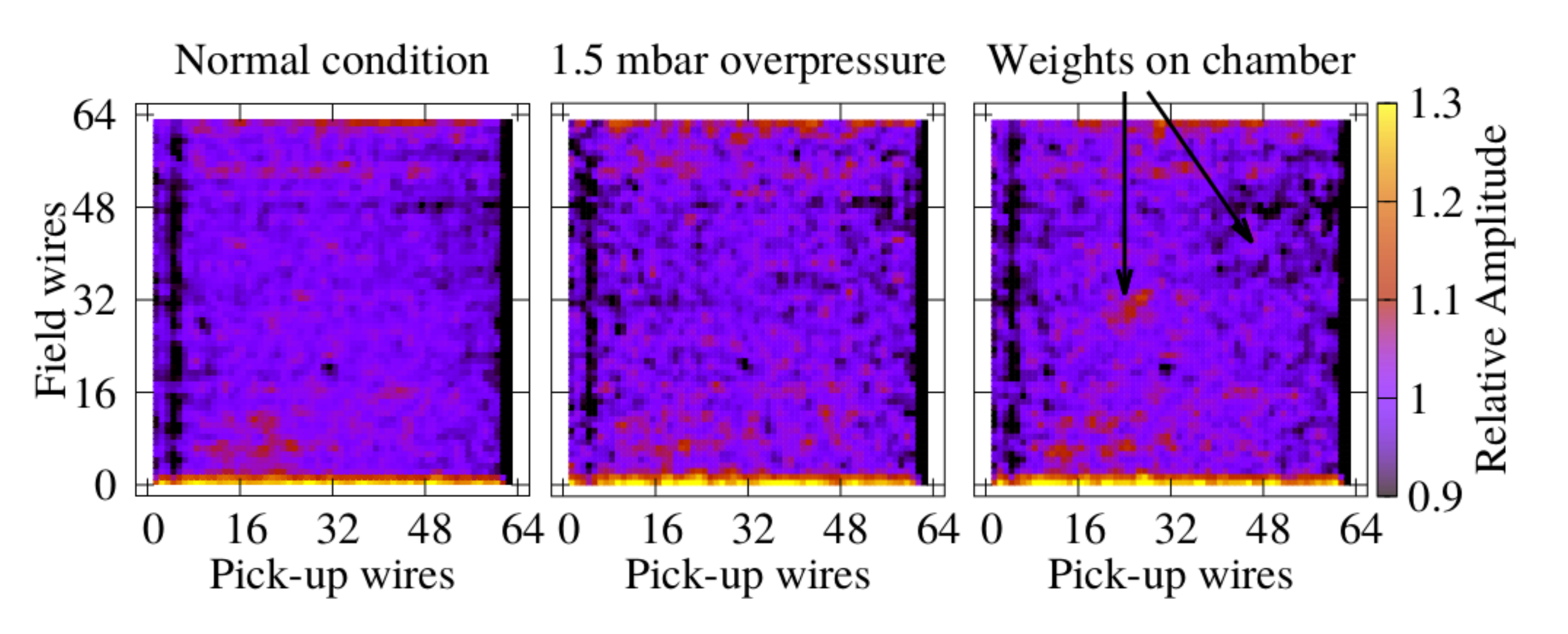}
\caption{(Color online) Gain map (mean amplitude) of the top detector (1) in the horizontal configuration, under various test conditions. The left panel is the reference, the middle panel shows the map for 1.5 mbar overpressure, the right panel with two test weights placed directly on the detector (indicated by the arrows).}
\label{fig:gain_uni}
\end{center}
\end{figure}

The gain map, that is, the mean signal amplitude over the detector surface under reference conditions is shown in the left panel of Figure~\ref{fig:gain_uni}. In the first study, the detectors were ran with an increased inner gas pressure of 1.5 mbar above the ambient atmosphere. This results in a slight bulging; the inner support pillars in this case act against the forces on the cathode layers. The corresponding gain map is shown in the middle panel of Figure~\ref{fig:gain_uni}, demonstrating that the gain drop under this condition is below 10\% (and hence, under normal operating conditions, one expects practically no efficiency loss).

The second study involved placing a weight directly on the detector. In this case, the detector was supported at the four corners, and two weights, 1kg each, was placed at the positions indicated in Figure \ref{fig:gain_uni}. One position was between the pillars, the other was approximately on one of the pillars. The gain increase (due to the reduction of the distance between the cathode layers) is apparent in the first case (by about 15\%), whereas at the position of the pillar, the change is smaller. There is no effect observable outside of the weight positions (e.g. no global detector bending). 

One can conclude that the mechanical structure, despite the simplicity, is very robust: the internal support pillars ensure the constant distance between cathode layers, and hence allow for acceptable low gain variation.

\section{Rate capability}
\label{sec:rate}

\begin{figure}[h!]
\begin{center}
\includegraphics[ angle=0, width=0.5\textwidth ]{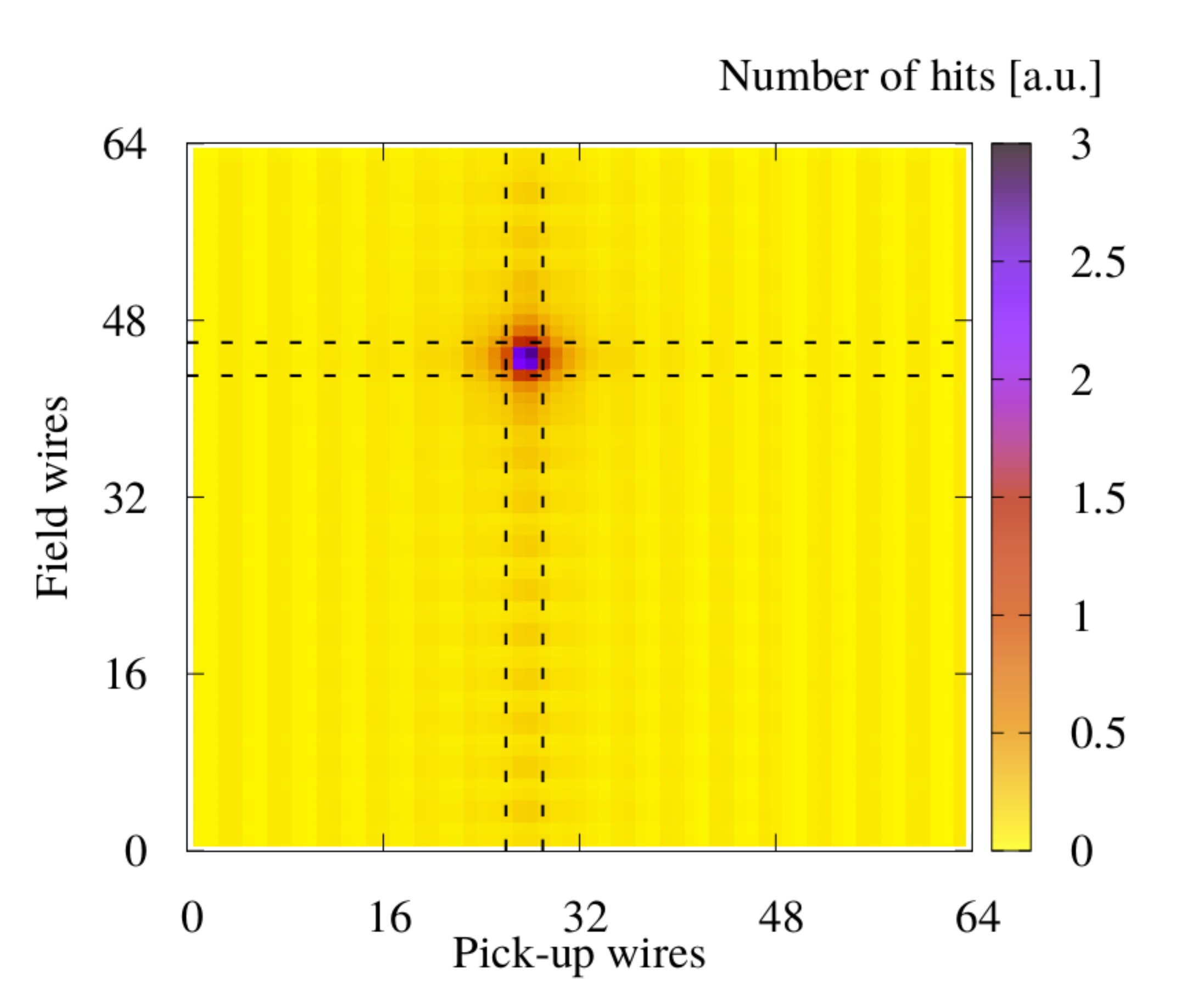}
\caption{(Color online) Distribution of all hits identified in the detector layer irradiated with the source. The local rate increase is apparent relative to the otherwise homogeneous distribution of cosmic-induced hits.}
\label{fig:source2d}
\end{center}
\end{figure}

In the standard setting, the tracking system has been operated in outdoor conditions on ground level, which is the maximum available cosmic muon rate. In order to quantify if a higher local rate is tolerable without efficiency loss, a $^{90}Sr$ beta source has been placed close to one of the layers. The native trigger count rate of the specific detector layer increased by 3.5 kHz (from around 1 kHz without source in horizontal position) from this concentrated irradiation covering about 20 ${\rm cm}^2$. Correspondingly, the number of hits appearing in triggered events (events in the sense of the tracking system) increases locally over the homogeneous distribution dominated by cosmic events, as shown in Fig.~\ref{fig:source2d}.

\begin{figure}[h!]
\begin{center}
\includegraphics[ angle=0, width=0.7\textwidth ]{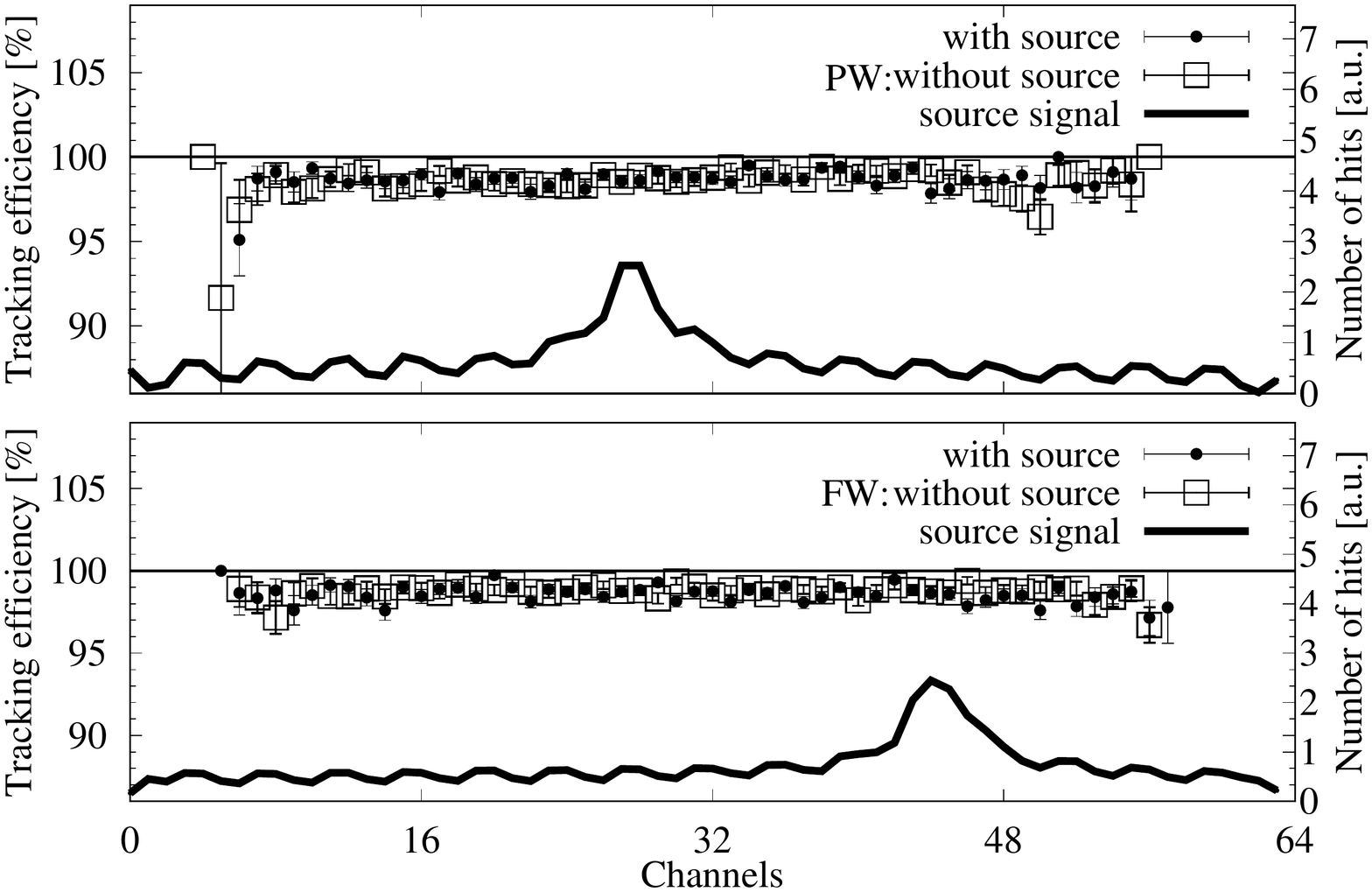}
\caption{Tracking efficiency determined along narrow strips overlapping with the source position. No indication of an efficiency loss (PW direction in the upper, FW direction in the lower panel) is observed, despite the considerable (uncorrelated) increase in hit number.}
\label{fig:source1d}
\end{center}
\end{figure}

The tracking efficiency has been determined along strips of 36 mm width (indicated by the dashed lines in Figure \ref{fig:source2d}) in both directions. The data was taken at an anode voltage of 1700 V, that is, at efficiency of 99.5 \%. %, to be sensitive to any local gain drop.
According to Figure \ref{fig:source1d}, there is no efficiency loss observed for the source region (shown with the peaked continuous line). 

The count rate by cosmic muons is in the order of 0.1 Hz/${\rm cm}^2$, whereas in our experiment the induced local rate is above 100 Hz/${\rm cm}^2$, corresponding to a three orders of magnitude increase. From the result, one can conclude that rate capability limitations, if any, are not relevant for the presented detector design in cosmic radiography applications.

\section{Required gas flow and tolerance on gas purity}
\label{sec:gasstudy}

The detector runs with industrial grade argon and carbon dioxide mixture in 82:18 proportion (Ferroline C18). The total volume of a single detector panel is around 13 liters, which implies long gas equilibration times for typical operational gas flow rates of 1-3 liters per hour.

In order to understand the behavior of the gas system, we have performed two studies. One of these involved direct injection of ambient air into the gas line, whereas the other was part of the realistic outdoor studies presented in Section \ref{sec:outdoor}. In the first case, the gas flow was set to 1 liter per hour, and the detectors were connected in series to the gas line (sequentially from 1 to 6). Using a gas bypass between detectors 1 and 2, 166 ${\rm cm}^3$ ambient air has been injected into the gas line, which then entered the second detector and later gradually moved along the subsequent panels. This way, the signal amplitude change relative to the first detector quantifies the effect of air (oxygen) contamination.

\begin{figure}[h!]
\begin{center}
\includegraphics[ angle=0, width=0.7\textwidth ]{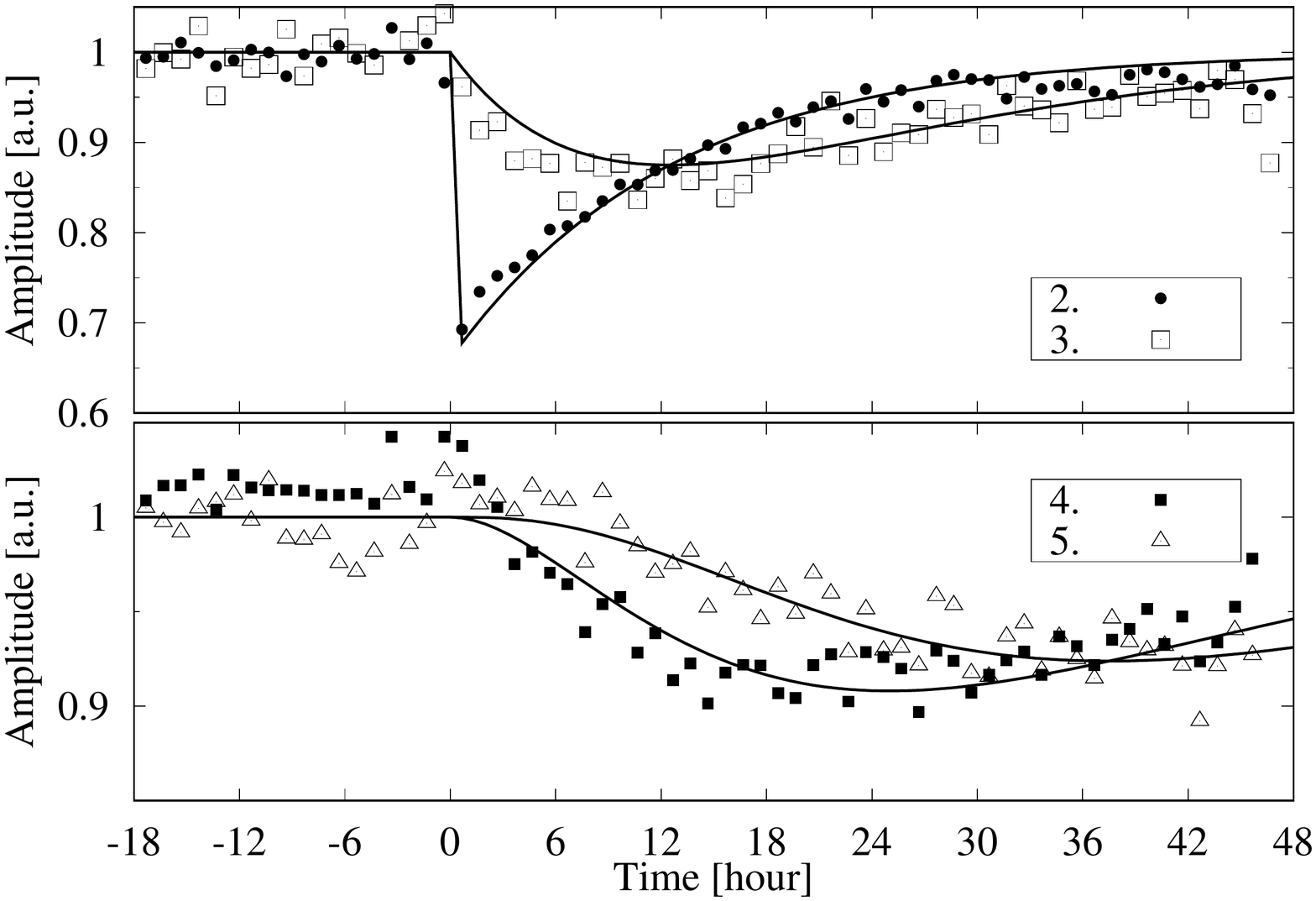}
\caption{Mean signal amplitude for detectors 2,3,4 and 5, relative to the first detector, after injection of ambient air to the gas flow, corresponding to 13000 ppm air contamination. The air was inserted from a bypass at $t=0$ into the second detector layer and flows along the gas line to the subsequent detectors. Predictions for the detectors, assuming full gas mixing (see the text for explanation) is indicated by the lines.}
\label{fig:gas_study}
\end{center}
\end{figure}

Figure \ref{fig:gas_study} shows the mean anode wire signal amplitude in the layers as a function of time $t$, normalized to the mean signal in detector 1 as reference before the gas injection (at $t=0$ in the Figure). The pattern is very clear, with a sharp decrease in the second detector, followed with an exponential purging.

Assuming complete mixing of the air bubble in the gas volume, indeed an exponential relaxation is expected for the mean signal amplitude $A$, described by

$$ \frac{\Delta A}{A} = -C e^{-t/T} $$

where the time constant $T$ is fixed by the ratio of the detector volume and the flow rate and found to be $T$= 12.5 hours. The only free parameter $C$, the relative gain drop, is determined to be $C$= 34 \%. The function is shown in Figure \ref{fig:gas_study} with the continuous line. In fact, the assumption of ``complete mixing'' predicts the gain drop in the subsequent layers, without any free parameters:

$$ \frac{\Delta A_k}{A_k} = -C \left( \frac{t}{T} \right)^{k-2} e^{-t/T}$$

for $k=2,3,4,5,6$, where $k$ is the number of the detector layer, and with $k=2$ we get back the pure exponential behavior. Figure \ref{fig:gas_study} confirms this behavior, indicated with the lines: the air bubble walks along the detectors.

The gain drop of 34\% corresponds to 1.3 \% (13000 ppm) air contamination; or assuming that only the oxygen content of air counts, 2500 ppm of $O_2$ contamination level. One can conclude that the gas distribution inside the detector system is fully predictable, with the fixed time constant $T$ determined as the detector volume filling time. Oxygen contamination of 1000 ppm, leading to less than 15\% gain drop (and hence no efficiency loss) thus seems to be tolerable, which is indeed a relaxed requirement for gas quality.

\section{Background and noise performance}
\label{sec:noise}

\begin{figure}[h!]
\begin{center}
\includegraphics[ angle=0, width=0.7\textwidth ]{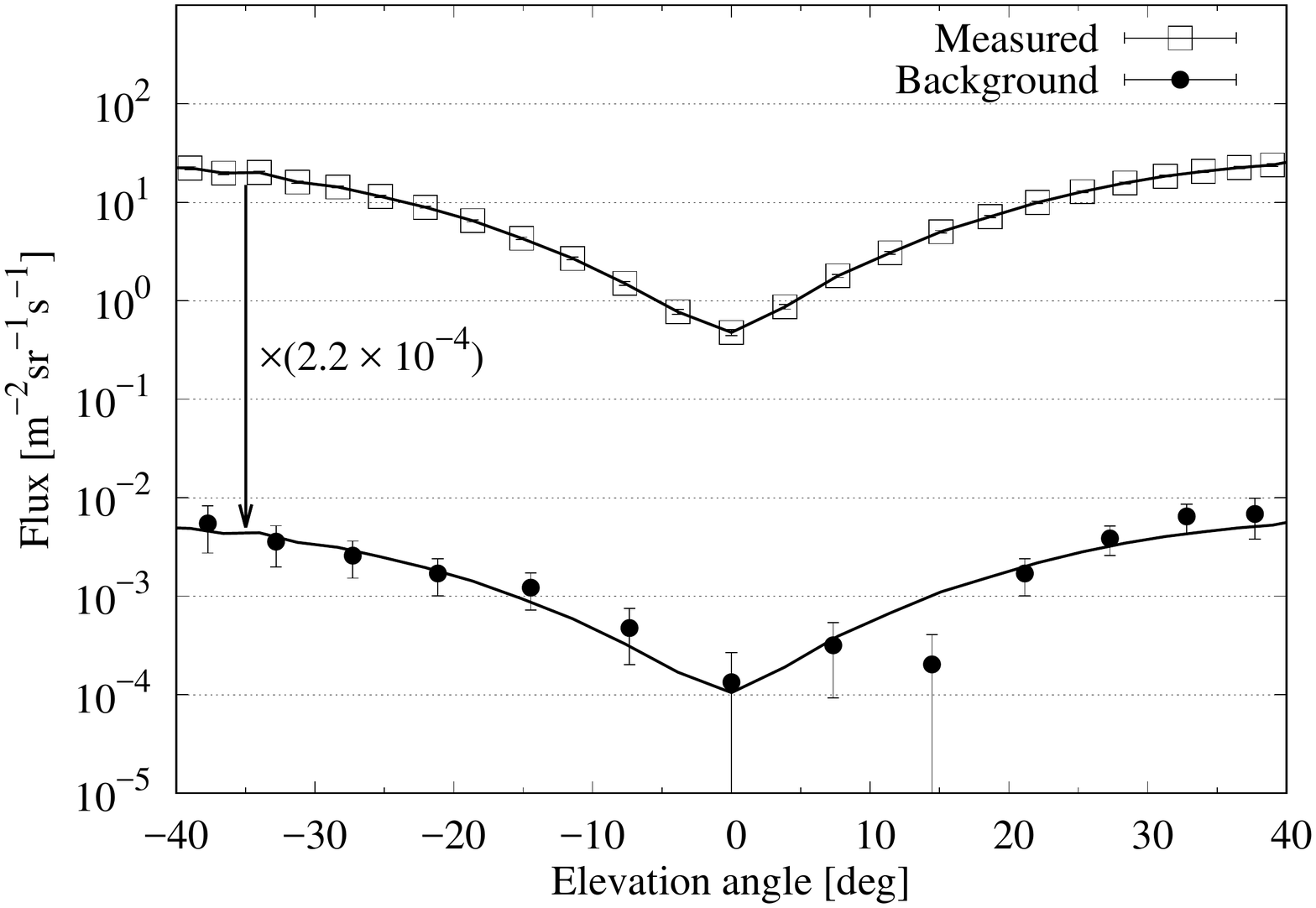}
\caption{The measured background flux $f_{BG}$, using random (oscillator) trigger. It is found to be nearly four orders of magnitude below the ``triggered'' particle flux $f$ (upper curve), and has a similar shape (indicated with the line) with a clear minimum at the horizon.}
\label{fig:flux_random}
\end{center}
\end{figure}

The instrumental background induced by the detector system is of vital importance from the point of view of practical muon radiography. Vaguely, we can call ``background'' anything which looks like a muon that traversed the object under study, while it actually did not. The most relevant component of such backgrounds is given by (low energy) particles which actually reach the detector~\cite{nishijama:2016}. This component can only be eliminated by a careful detector design and will be subject of future detailed studies; however, at this point we aim at quantifying instrumental artifacts (background which may not even be connected to any particle depositing charge in the detector). One possibility is to apply a ``random'' (uncorrelated oscillator) trigger and observe patterns resembling trajectories. In the first place, it has been verified that the pure ``electronics'' noise is exceedingly small: with the high voltage off, simply there are no triggers, and there are no observable trajectories. The probability of any of the single channels (on FW or PW) firing is in the order of $10^{-6}$, that is, the chance to have 5 or 6 of them in a line to fake a track is negligible. In order to have a more relevant background estimate, the tracking system was operated in a fully efficient configuration ($U_{SW}=1700V$) with vertical detectors, and was running for a day with oscillator trigger at 110 Hz. This data contains reconstructed tracks, and allows to estimate the ``background'' flux $f_{BG}$. The true particle flux $f$ can be calculated with standard procedures~\cite{mt_nima:2012}, and is proportional to the number of observed tracks per event:

$$f \propto \frac{N}{t} = \frac{N}{N_{trig}} \frac{N_{trig}}{t} = \frac{N}{N_{trig}} R_{trig}$$

where the number of tracks $N$ are measured over time of $t$, whereas the number of triggered events $N_{trig}$ divided by $t$ is the trigger rate $R_{trig}$. $N/N_{trig}$ is the mean track number per event. Most triggered events contain only short tracks (due to the weak trigger condition), therefore the background tracks $N_{BG}$ add up to the ``true'' track set. The probability of random tracks, per event (taken with the oscillator trigger), can be used to estimate the background flux $f_{BG}$: 

$$f_{BG} \propto \frac{N_{BG}}{N_{osc}} R_{trig}$$

where $N_{osc}$ is the event number taken with the oscillator (random) trigger. The background flux $f_{BG}$ has been evaluated as a function of elevation angle and is shown in Figure \ref{fig:flux_random}. There is a clear minimum at the horizon (elevation angle 0), reaching a background flux value of around $2 \times 10^{-4} m^{-2} sr^{-1} s^{-1}$. The background flux happens to be around 4 orders of magnitude below the true measured flux $f$ (also indicated in the Figure). One has to note that this measured flux has nothing to do with the flux of cosmic muons (which should reduce to far lower values near the horizon): the measurement includes a considerable and undetermined low energy background.

%Most of these background tracks are expected to be randomly coincident true particles, which is suggested by the similarity of the shape of the two fluxes. One can actually estimate the time frame $\Delta t$ during which random particle tracks are observable in the detector. The probability to find a particle track in a randomly triggered event is the same as the probability to be within $\Delta t$ time to a true particle track during the observation time $t$:

% $$\frac{N_{BG}}{N_{trig}} = \frac{N}{t} \Delta t = \frac{N}{N_{trig}} R_{trig} \Delta t$$

% Considering that in our case $R_{trig}$ was 45 Hz, whereas the background over true flux ratio is around $2.2 \times 10^{-4}$, one gets approximately $\Delta t$=10 $\mu$s, a number fairly consistent with the expectation from the timing scan presented in Section \ref{sec:efficiency}.

The key conclusion of the observations above is that there is no apparent uncorrelated track background beyond that expected from the truly random appearance of otherwise physical tracks; such tracks are expected to be suppressed along with the low energy background component in true muography tracking systems.

\section{Experience under ambient outdoor conditions}
\label{sec:outdoor}

\begin{figure}[h!]
\begin{center}
\includegraphics[ angle=0, width=0.8\textwidth ]{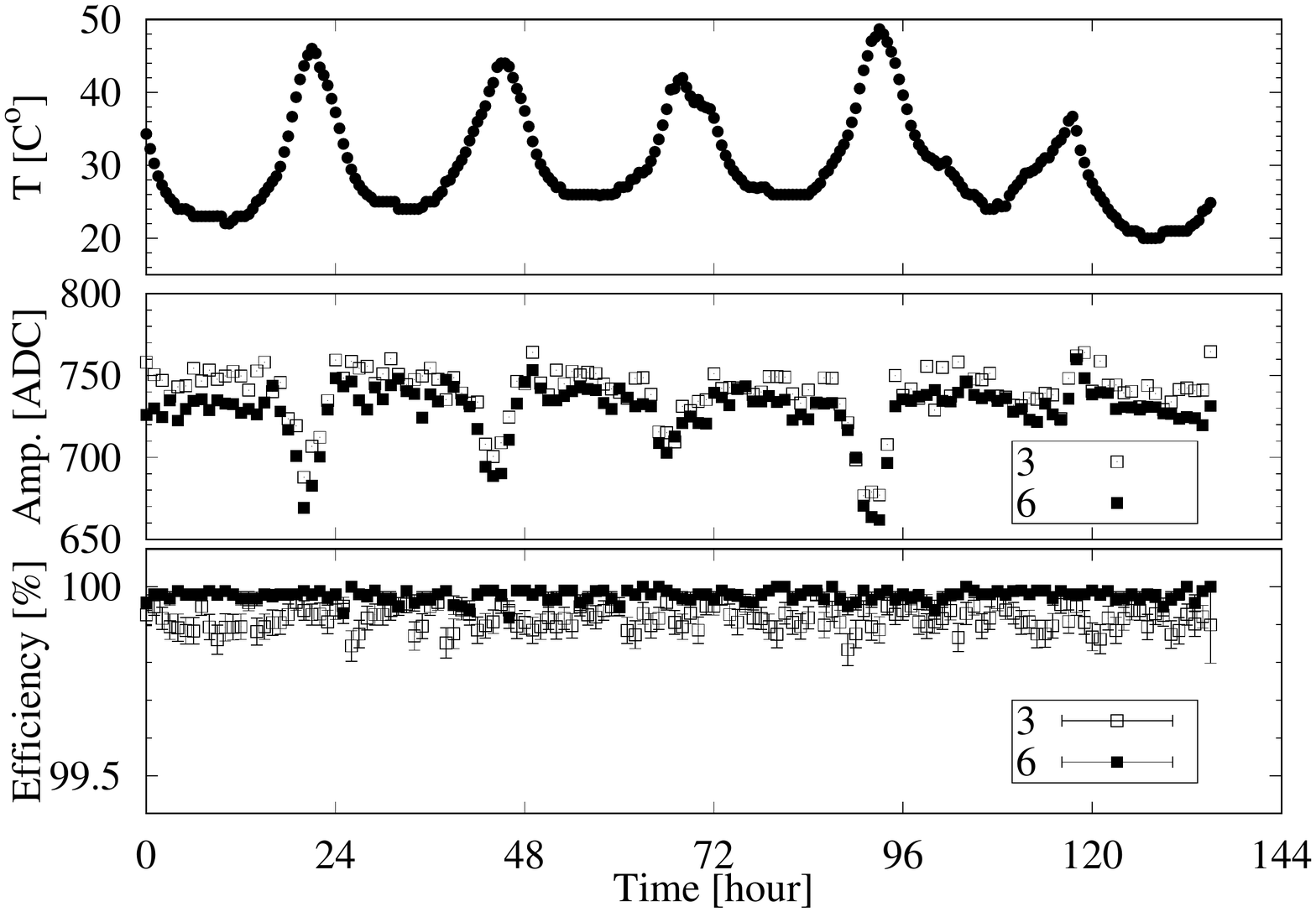}
\caption{Time dependence of relevant operational parameters during 6 days of data taking in outdoor conditions. Temperature variations up to 25 $^{\circ}$C has been sustained (top panel), with not more than 10\% mean amplitude changes (due to partial compensation of the anode HV). The tracking efficiency has been consistently high, shown in the lower panel, during the whole period (note the scale on the lower panel starts at 99.5\%).}
\label{fig:stability}
\end{center}
\end{figure}

For testing under ambient outdoor conditions, the tracking system with vertical detectors have been installed on a support frame and placed on the flat roof of a building at the Wigner Research Centre campus. To allow for considerable environmental influence, the system was only lightly covered with a box, thus receiving strong sunlight in the early afternoon. In Figure \ref{fig:stability}, the outside temperature is shown on the top panel, featuring more than 25 $^{\circ}$C daily temperature variations. During the 6 days of continuous data taking in our example, the system sustained a light raining on the 5th day, otherwise the weather was clear.

The change of the temperature by 30 $^{\circ}$C corresponds to nearly 10\% change in absolute temperature, and hence in gas density. The system thus can not run at constant anode voltage, but needs to be compensated to first order: in our system, the anode HV was reduced by 1 V for a 1 $^{\circ}$C temperature increase (from the nominal 1700 V at 25 $^{\circ}$C). Though the compensation is far from perfect (possibly containing non linear terms), it has been verified that this ensures only a slight, 10\% mean amplitude variation (demonstrated on the middle panel).

The key figure of merit for the detector is the tracking efficiency, shown in the lower panel of Figure \ref{fig:stability}. The measured tracking efficiency is reliably high ($>$99.5\%) and stable for the whole period (shown only for two detector layers but found consistently for all panels).

The detector was running with a constant, 2 liters per hour gas flow rate. This presents an interesting issue during decreasing temperatures. The total tracking system gas volume was about 100 liters. Assuming that the temperature decreases by 6 $^{\circ}$C in one hour, the 2\% relative change in absolute temperature causes the total gas volume to reduce accordingly, that is, by 2 liters. In this case, actually air may be sucked in from the outside atmosphere to the last detector. In order to avoid this, a long buffer tube has been added after the last detector, with sufficiently large (in our case 1 liters) total volume.

\section{Summary and conclusions}
\noindent

A new approach to the classical MWPC-s has led to a highly efficient, low weight, mechanically and environmentally tolerant tracking detector, which by its simple and cost efficient construction, can be an ideal part of large size muon radiography detectors. An internal support pillar structure has been shown to present a negligible dead zone. The tracking system has been demonstrated to tolerate broad temperature variations maintaining very high detection efficiency ($>$99.5\%). The position resolution of better than 4 mm RMS (9mm FWHM) is compatible with the segmentation. Gas flow rate using non-flammable, non-toxic Ar and CO$_2$ mixture of 1-2 liters per hour is sufficient for the operation, with a well predictable gas flow pattern. Future complete muography tracking systems will be completed with absorber (scatterer) layers and are intended to be applied for large size object imaging, such as the internal structure of volcanos.

%%%%%%%%%%%%%%%%%%%%%%%%%%%%%%%%%%%%%%%%%%%%%%%%%%%%%%%%%%%%%%%%%%%%%%%%%%%%%%%%%%%%%%%%%%%%%%%

\section*{Acknowledgement}

\label{sec:ack}

\noindent
Support was provided by members of the REGARD group of the Wigner Research Centre for Physics of the Hungarian Academy of Sciences. We are grateful to Prof. H. K. M. Tanaka for the useful discussions. This work was supported by the Momentum (Lend\"ulet) grant of the Hungarian Academy of Sciences under the grant number LP2013-60.

%%%%%%%%%%%%%%%%%%%%%%%%%%%%%%%%%%%%%%%%%%%%%%%%%%%%%%%%%%%%%%%%%%%%%%%%%%%%%%%%%%%%%%%%%%%%%




%%% Standard TheBibliography

\end{document}